\documentclass[twocolumn,showpacs,showkeys,preprintnumbers,amsmath,amssymb,superscriptaddress]{revtex4-1}
\usepackage{graphicx}
\usepackage{natbib}
\usepackage{color}
\usepackage{hyperref}
\hypersetup{
 colorlinks,
 linkcolor={red},
 citecolor={blue},
 urlcolor={blue}
}
\hypersetup{breaklinks=true}

\begin{document}
\typeout{Paper in revtex 4.1i 2009/10/19 (AO)}

\title{First direct measurement of isotopic fission-fragment yields of $^{239}$U}
\author{D.~Ramos}
\email[Email address: ]{diego.ramos@ganil.fr}
\affiliation{IPN Orsay, Universit\'e de Paris-Saclay, CNRS/IN2P3, F-91406 Orsay Cedex, France}
\affiliation{GANIL, CEA/DRF-CNRS/IN2P3, BP 55027, F-14076 Caen Cedex 5, France}
\author{M.~Caama\~no}
\affiliation{IGFAE, Universidade de Santiago de Compostela, E-15706 Santiago de Compostela, Spain}
\author{A.~Lemasson}
\affiliation{GANIL, CEA/DRF-CNRS/IN2P3, BP 55027, F-14076 Caen Cedex 5, France}
\author{M.~Rejmund}
\affiliation{GANIL, CEA/DRF-CNRS/IN2P3, BP 55027, F-14076 Caen Cedex 5, France}
\author{L.~Audouin}
\affiliation{IPN Orsay, Universit\'e de Paris-Saclay, CNRS/IN2P3, F-91406 Orsay Cedex, France}
\author{H.~\'Alvarez-Pol}
\affiliation{IGFAE, Universidade de Santiago de Compostela, E-15706 Santiago de Compostela, Spain}
\author{J.D.~Frankland}
\affiliation{GANIL, CEA/DRF-CNRS/IN2P3, BP 55027, F-14076 Caen Cedex 5, France}
\author{B.~Fern\'andez-Dom\'inguez}
\affiliation{IGFAE, Universidade de Santiago de Compostela, E-15706 Santiago de Compostela, Spain}
\author{E.~Galiana-Bald\'o}
\affiliation{IGFAE, Universidade de Santiago de Compostela, E-15706 Santiago de Compostela, Spain}
\affiliation{LIP Lisboa, 1649-003 Lisbon, Portugal}
\author{J.~Piot}
\affiliation{GANIL, CEA/DRF-CNRS/IN2P3, BP 55027, F-14076 Caen Cedex 5, France}
\author{D.~Ackermann}
\affiliation{GANIL, CEA/DRF-CNRS/IN2P3, BP 55027, F-14076 Caen Cedex 5, France}
\author{S.~Biswas}
\affiliation{GANIL, CEA/DRF-CNRS/IN2P3, BP 55027, F-14076 Caen Cedex 5, France}
\author{E.~Clement}
\affiliation{GANIL, CEA/DRF-CNRS/IN2P3, BP 55027, F-14076 Caen Cedex 5, France}
\author{D.~Durand}
\affiliation{LPC Caen, Universit\'e de Caen Basse-Normandie-ENSICAEN-CNRS/IN2P3, F-14050 Caen Cedex, France}
\author{F.~Farget}
\affiliation{LPC Caen, Universit\'e de Caen Basse-Normandie-ENSICAEN-CNRS/IN2P3, F-14050 Caen Cedex, France}
\author{M.O.~Fregeau}
\affiliation{GANIL, CEA/DRF-CNRS/IN2P3, BP 55027, F-14076 Caen Cedex 5, France}
\author{D.~Galaviz}
\affiliation{LIP Lisboa, 1649-003 Lisbon, Portugal}
\author{A.~Heinz}
\affiliation{Chalmers University of Technology, SE-41296 G\"oteborg, Sweden}
\author{A.I.~Henriques}
\affiliation{CENBG, IN2P3/CNRS-Universit\'e de Bordeaux, F-33175 Gradignan Cedex, France}
\author{B.~Jacquot}
\affiliation{GANIL, CEA/DRF-CNRS/IN2P3, BP 55027, F-14076 Caen Cedex 5, France}
\author{B.~Jurado}
\affiliation{CENBG, IN2P3/CNRS-Universit\'e de Bordeaux, F-33175 Gradignan Cedex, France}
\author{Y.H.~Kim}
\thanks{Present address: ILL, F-38042 Grenoble Cedex 9, France}
\affiliation{GANIL, CEA/DRF-CNRS/IN2P3, BP 55027, F-14076 Caen Cedex 5, France}
\author{P.~Morfouace}
\thanks{Present address: CEA, DAM, DIF, F-91297 Arpajon, France}
\affiliation{GANIL, CEA/DRF-CNRS/IN2P3, BP 55027, F-14076 Caen Cedex 5, France}
\author{D.~Ralet}
\affiliation{CSNSM, CNRS/IN2P3, Universit\'e de Paris-Saclay,F-91405 Orsay, France}
\author{T.~Roger}
\affiliation{GANIL, CEA/DRF-CNRS/IN2P3, BP 55027, F-14076 Caen Cedex 5, France}
\author{C.~Schmitt}
\affiliation{IPHC Strasbourg, Universit\'e de Strasbourg-CNRS/IN2P3, F-67037 Strasbourg Cedex 2, France}
\author{P.~Teubig}
\affiliation{LIP Lisboa, 1649-003 Lisbon, Portugal}
\author{I.~Tsekhanovich}
\affiliation{CENBG, IN2P3/CNRS-Universit\'e de Bordeaux, F-33175 Gradignan Cedex, France}
\date{\today}

\begin{abstract}

A direct and complete measurement of isotopic fission-fragment yields of $^{239}$U has been performed for the first time. The $^{239}$U fissioning system was produced with an average excitation energy of 8.3~MeV in one-neutron transfer reactions between a $^{238}$U beam and a $^{9}$Be target at Coulomb barrier energies. The fission fragments were detected and isotopically identified using the VAMOS++ spectrometer at the GANIL facility. This measurement allows to directly evaluate the fission models at excitation energies of fast neutrons, relevant for next-generation nuclear reactors. The present data, in agreement with model calculations, do not support the recently reported anomaly in the fission-fragment yields of $^{239}$U and confirm the persistence of spherical shell effects in the Sn region at excitation energies exceeding the fission barrier by few MeV. 

\end{abstract}


\pacs {}

\maketitle

Eighty years after its discovery~\cite{Meitner1939,Hahn1939}, fission continues to play a major role in the production of electricity~\cite{NEA-report}, and it is a key process for the management and the transmutation of long-lived radioactive nuclear waste~\cite{transmutation}. Fissioning systems also serve as natural laboratories to study nuclear dynamics~\cite{Hamilton1995,Wada1993,Lazarev1993,Jurado2004}, are tools to produce neutron-rich nuclei and study their structure~\cite{Hamilton1995,Adrich05}, and play a role in the r-process nucleosynthesis~\cite{Goriely2013}. However, a complete microscopic quantum description of the fission process is still lacking~\cite{Schunck2016}. At low excitation energy, the fission mechanism is particularly challenging because of the complex interplay of dynamic and static properties that drives the fissioning system to fission fragments~\cite{Schmidt2018,Andreyev2018}. This includes nuclear configurations far from equilibrium, the interplay of collective and intrinsic degrees of freedom, and the dynamics of large amplitude collective motion~\cite{Bulgac2016,Scamps2018,wil76,Bro90,Moller2001}.

The description of the fission process strongly relies on available experimental information obtained from the final fission fragments~\cite{Andreyev2018}. The key observables are fission yields, kinetic energies, and deexcitation schemes of fission fragments. Until recently, the access to these observables was limited to neutron-induced fission on long-lived or stable nuclei. The complete fission-fragment identification was not feasible due to the low kinetic energy of the fission products. The use of surrogate reactions gave access to the study of a wider range of compound nuclei, otherwise inaccessible~\cite{Jutta2012,Kessedjian2010,Hirose2017}. The use of inverse kinematics allowed the direct measurement of the atomic number of complete fission-fragment distributions~\cite{Schmidt2000}. In the last decade, the simultaneous use of surrogate reactions, inverse kinematics, and magnetic spectrometers has opened a new field of study measuring complete isotopic fission-fragment distributions~\cite{Caamano2013a,JLRS2015,Pellereau2017,Ramos2018}, and leading to an improved understanding of the fission process~\cite{Schmidt2018}. 

There is a need of accurate experimental information on fission fragments because the state-of-art evaluated data for many systems at fast-neutron energies~\cite{JEFF,ENDF} rely on interpolations and empirical models. This is required, in particular, for the modeling of next generation reactors, such as future Fast Reactors (FR) and Accelerator-Driven Systems (ADS)~\cite{FR-ADS}. The above-mentioned surrogate reactions at Coulomb barrier energies involving transfer-induced fission~\cite{Caamano2013a,Ramos2018} represent unique opportunities to collect such relevant data.
 
In this letter, the first direct measurement of the isotopic fission-fragment yields of $^{239}$U is reported. These results confirm the persistence of the fission path characterized by a heavy fragment at low deformation (\textit{Standard I mode}~\cite{Bro90}) at fast-neutron energies. Recently, anomalies in the fission-fragment yields were reported for neutron induced fission of $^{238}$U~\cite{Wilson2017}, with large deviations (up to 600\%) with respect to models. These anomalies would change the evaluation of the heat from fission-fragment $\gamma$-decays, which is necessary for modeling present and future reactors. For instance, $^{238}$U(n,f) reactions contribute with 2\% to 5\% of the total fission rate in current Pressurized Water Reactors (PWR)~\cite{J.Duderstadt,Marguet2017} and they compete with neutron-capture reactions in certain future fast-reactor designs.
The present work rules out the reported anomalies on the fission-fragment yields of Mo and Sn and gives reliable constraints for current fission models at fast-neutron energies.


The experiment was performed at GANIL using a beam of $^{238}$U at 5.88~AMeV impinging on a $500~\mu$g/cm$^{2}$-thick $^{9}$Be target. $^{239}$U was produced in flight in one-neutron transfer reactions, $^{9}$Be$(^{238}$U,$^{239}$U$)^{8}$Be, with a range of excitation energies high enough to overcome the fission barrier ($B_f=6.4$~MeV~\cite{Bj1980}) and undergo fission. The fission fragments were detected in the VAMOS++ magnetic spectrometer~\cite{Rejmund2011} in coincidence with two $\alpha$ particles, resulting from the breakup of $^{8}$Be, detected in the SPIDER telescope~\cite{Carme2014}, placed at 31 mm downstream from the target.
The segmentation of SPIDER provides a measurement of the angle of the recoil from $35^{\circ}$ to $55^{\circ}$ with respect to the beam axis. This, combined with the measurement of the total energy, allows an event-by-event determination of the total excitation energy of the system with a typical resolution of $1.7$~MeV. Due to the kinematical focussing, fission fragments were emitted at forward angles, within a cone of $\sim 30^{\circ}$.
For each measured fission event, one of the two fragments was fully characterized in terms of the mass number, atomic number, atomic charge, and velocity vector using the VAMOS++ spectrometer and its associated detectors.
The VAMOS++ spectrometer was rotated by $14^{\circ}$ and $21.5^{\circ}$ with respect to the beam axis to optimize the acceptance of heavy and light fragments, respectively. 
Further details on  VAMOS++ along with typical performances for the fission-fragments detection are given in Refs.~\cite{Rejmund2011,Kim2017,Vandebrouck2016}. 

\begin{figure}[t!]
\includegraphics[width=0.5\textwidth]{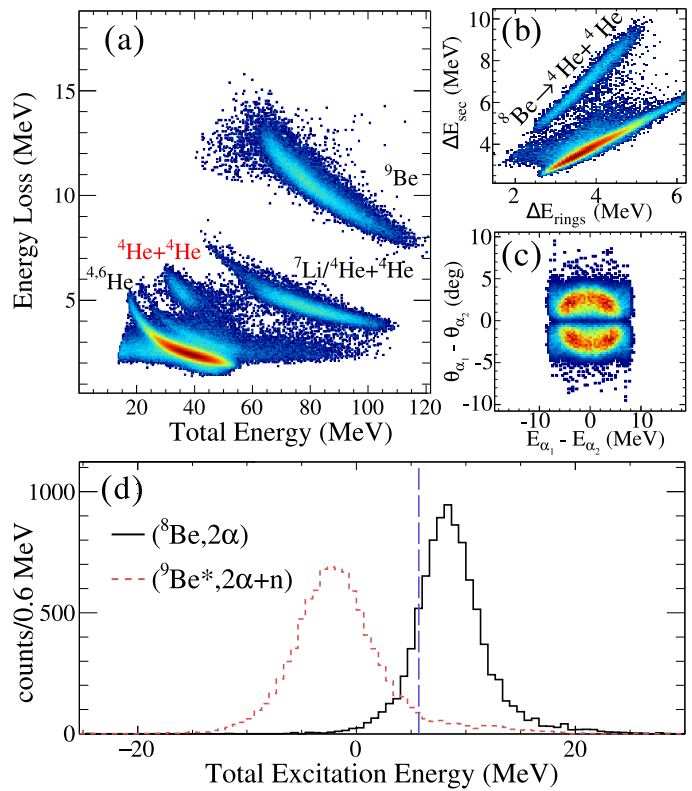}
\caption{(Color online) Identification of target-like ions detected in coincidence with fission fragments and total excitation energy of the fissioning system:
(a) Correlation between the energy loss and the total energy for the target-like ions detected in the SPIDER telescope. (b) Correlation of the energy loss measured in the rings and the corresponding sector. (c) Correlation between the polar angle and the energy difference of two coincident $\alpha$ particles. (d) Reconstructed total excitation energy of the fissioning system, in coincidence with fission events, assuming either the breakup of $^{8}$Be (solid line) or $^{9}$Be (dotted line). The vertical dashed line represents the fission barrier of $^{238}$U (see text for details).}
\label{fig::Identification}
\end{figure}

\begin{figure*}[t!]
\includegraphics[width=1\textwidth]{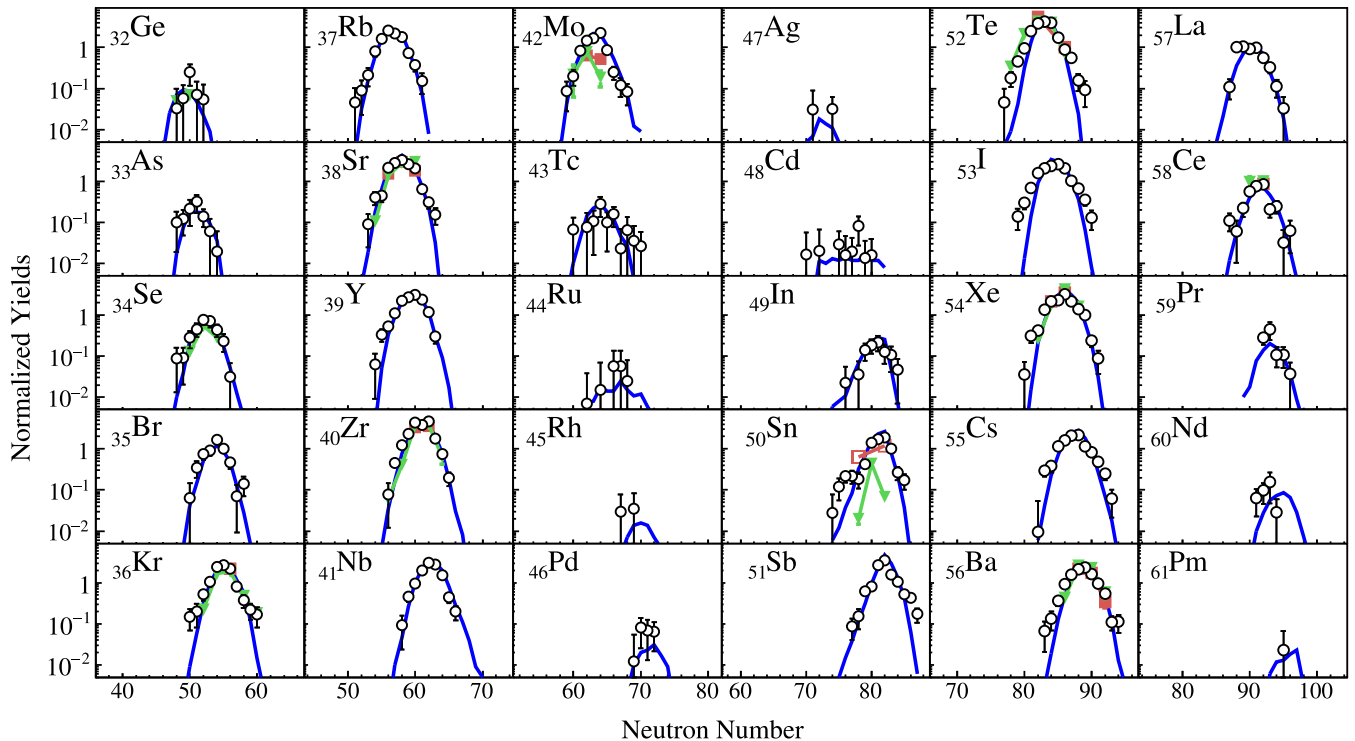}
\caption{
(Color online) Isotopic fission yields of $^{239}$U (higher than $10^{-2}$) : Each panel corresponds to an element, ranging from Ge to Pm. Yields are shown as a function of the neutron number. The present data (open circles) is compared with the results of GEF (2018/1.1)~\cite{GEF} (blue lines) and with the results of $\gamma$-spectroscopy measurements in Ref.~\cite{Wilson2017} (green triangles and lines) and Ref.~\cite{Fotiades2019} (red squares and lines). \label{fig::IsotopicYields}}
\end{figure*}

Figure~\ref{fig::Identification}(a) shows the identification spectrum of target-like ions, detected in coincidence with fission fragments, obtained from the correlation of the energy loss and the total energy, measured in SPIDER.  The energy loss of the two coincident $\alpha$ particles, detected in the same sector, is similar to the energy loss of $^{7}$Li. Therefore, the selection of actual $\alpha$-$\alpha$ coincidences is obtained by selecting events where each $\alpha$ particle hits a different ring within the same sector. For such events, the energy loss in the sector is twice the energy loss in each ring. This is shown in Fig.~\ref{fig::Identification}(b), where the $\alpha$-$\alpha$ coincidence was not applied. At the edge of the detector, one $\alpha$ particle may escape from the telescope without hitting the second detector. In such a case, the measurement of the energy is incomplete and those events were discarded (red label in Fig.~\ref{fig::Identification}(a)). Figure~\ref{fig::Identification}(c) shows the correlation between the polar angle and the energy difference of coincident $\alpha$ particles that corresponds to two-body decay. In Fig.~\ref{fig::Identification}(d), the reconstructed total excitation energy distribution, obtained assuming that the two coincident $\alpha$ particles follow the breakup of $^{8}$Be is shown with a solid line, for those events detected in coincidence with fission fragments. 

These events can contain a small fraction corresponding to the breakup of $^{9}$Be from its unbound first excited state at $1.684$~MeV~\cite{Macdonald1992}. The dotted line of Fig.~\ref{fig::Identification}(d) shows the distribution assuming the breakup of $^{9}$Be. It can be seen that in the case of $^{9}$Be breakup, only the small fraction of the excitation energy distribution above the fission barrier of $^{238}$U ($B_f = 5.7$~MeV~\cite{Bj1980}), indicated by the vertical dashed line, can undergo fission. These events, with excitation energies above 15~MeV assuming $^{8}$Be breakup, were not considered in the following analysis to avoid contamination. Additional contamination from random coincidences between the $^{8}$Be and fission fragments from fusion-fission reactions was observed to be $(4\pm0.5)\;\%$ of the measured neutron-transfer induced fission. This was isotopically subtracted following the procedure described in Ref.~\cite{Ramos2018}. This analysis procedure ensures that the contribution of other fissioning systems is lower than 0.9\%.

Isotopic-fission yields were derived following the procedure presented in Refs.~\cite{Caamano2013a, Ramos2018} within the range of excitation energy $0\leq E_x\leq 15$ MeV, resulting in a mean excitation energy of $8.3$~MeV with a standard deviation of $2.7$~MeV.
The total uncertainties presented in the data are obtained as the quadratic sum of statistical and systematic uncertainties. The systematic uncertainties, ranging from 2\% in the heavier fragments up to 10\% in the lighter ones, include those from the determination of the spectrometer acceptance, the relative normalization between both settings, and the contamination subtraction from fusion-fission. 


\begin{figure}[t!]
\includegraphics[width=0.5\textwidth]{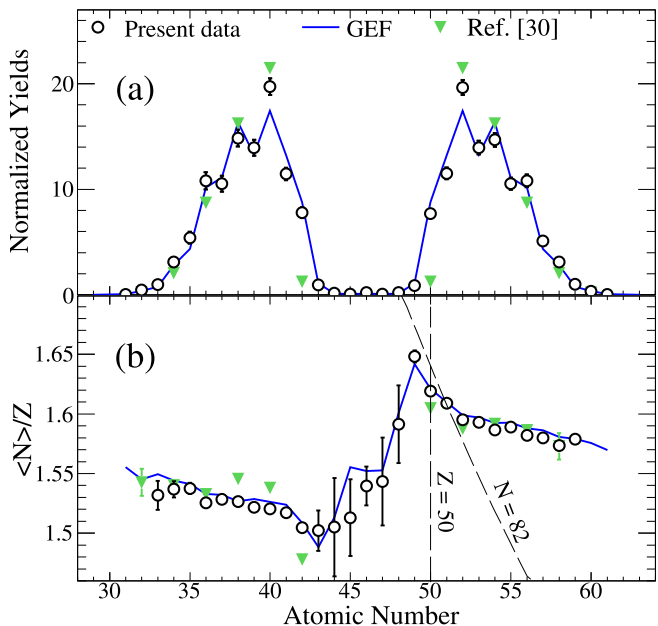}
\caption{(Color online) Elemental fission yields and average neutron excess of fission fragments: (a) Normalized yield as a function of the atomic number. Present data (open circles) is compared with the results of GEF (2018/1.1)~\cite{GEF} (blue line) and with the data from $\gamma$-spectroscopy~\cite{Wilson2017} (green triangles). (b) Average neutron excess of fission fragments as a function of the atomic number. Spherical closed shells $Z=50$ and $N=82$ are indicated by dashed lines. \label{fig::ElementalYields}}
\end{figure}

Isotopic fission yields of $^{239}$U (open circles) are presented in Fig.~\ref{fig::IsotopicYields} as a function of the fragment neutron number. These are compared with the semi-empirical model~\textit{General Description of Fission Observables} (GEF)~\cite{GEF}, commonly used in nuclear data evaluation. The GEF calculation (lines) was obtained using the measured excitation energy distribution assuming a transfered angular momentum of $J=3\;\hbar$ (the sensitivity to this parameter is discussed later on). The results of $\gamma$-spectroscopy measurements of neutron-induced fission at $E_x=6.5$~MeV from Ref.~\cite{Wilson2017} (green triangles) and Ref.~\cite{Fotiades2019} (red squares) are also shown. The yields for the intermediate elements are dominated by statistical fluctuations due to the strong asymmetry of the fission of $^{239}$U, with a very low production of Rh, Pd, and Ag. A good agreement between the present data and the GEF calculations is obtained for the light fission products, while for the heavy fission products, the width of the distributions are underestimated. A satisfactory agreement is also found between the present data and those obtained from $\gamma$-spectroscopy, except for Mo and Sn isotopes. In Sn isotopes, the data from Ref.~\cite{Fotiades2019} corresponds to upper limits (unfilled red squares).
The distributions of Mo and Sn, complementary isotopes, show mirror asymmetric shapes in both the present data and GEF calculations. These shapes can be described in terms of different fission modes as discussed in Ref.~\cite{Pellereau2017}. 

Fission yields of $^{239}$U are shown in Fig.~\ref{fig::ElementalYields}(a), as a function of the atomic number of the fission fragment. The present data (open circles) is compared with the GEF calculation (lines) and with the data obtained in the previous $\gamma$-spectroscopy measurements~\cite{Wilson2017} (triangles). The present data shows a strong even-odd effect that is fairly well reproduced by the GEF calculation. The asymmetric fission is also well reproduced by GEF, with an agreement within $10\;\%$, except for Te and Zr, which are underestimated. A good agreement is found between both sets of experimental data for most of the elements within $\sim 20\;\%$. However, large differences are observed in Mo and Sn where the present dataset do not show the deviation by $600\;\%$ with respect to the models reported in Ref.~\cite{Wilson2017}.

Figure~\ref{fig::ElementalYields}(b) shows the average neutron excess for the fission fragments after neutron evaporation, defined as the average number of neutrons of each element divided by its atomic number, as a function of the atomic number. The dashed lines correspond to spherical closed shells $N=82$ and $Z=50$. The effect of these closed shells is clearly observed in the neutron excess of the heavy fragments, where the amount of neutrons of the fragments is locally enhanced, due to the double-magicity of $^{132}$Sn. The present data and the GEF calculations exhibit a very good agreement in the region of Sn. 
The data from $\gamma$-spectroscopy shows significant fluctuations. Both Mo and Sn show a clear reduction of their mean neutron excess with respect to the present data. This is in contradiction with the expected increase the neutron evaporation, mainly for the heavy fragment~\cite{Naqvi1986} due to the additional excitation energy of present data (+2 MeV). Consequently, the neutron excess should be reduced compared to the $\gamma$-spectroscopy data. This opposite behavior suggests an experimental bias in the measured isotopic distribution from $\gamma$-spectroscopy for neutron-rich isotopes. 

\begin{figure}[t!]
\includegraphics[width=0.5\textwidth]{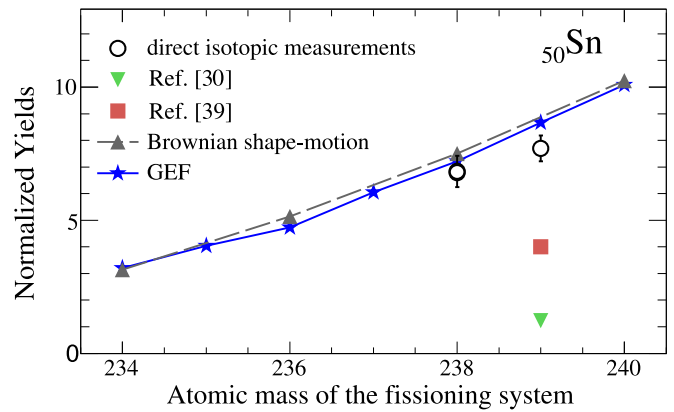}
\caption{(Color online) Fission yields for Sn: Evolution of the yields for Sn as a function of the atomic mass of the fissioning system for the isotopic chain of uranium. The results of the direct isotopic measurements: obtained in this work and from Refs.~\cite{Ramos2018,Pellereau2017} (open circles) are compared with a model based on the Brownian shape-motion and on 5D potential-energy surfaces~\cite{Moller2017} (dashed line) and with the GEF code (2018/1.1)~\cite{GEF} (solid line). Data from ${\gamma}$-spectroscopy measurement Ref.~\cite{Wilson2017} (triangle) and Ref.~\cite{Fotiades2019} (square) are also included, in the last case obtained from the complementary fragment. \label{fig:SnYield}} 
\end{figure}

Figure~\ref{fig:SnYield} shows the evolution of the yield for Sn fragments from fission of uranium as a function of the atomic mass of the fissioning system. The Brownian shape-motion model with random walks on 5D potential-energy surfaces~\cite{Moller2017} predicts a continuous increase of the yield for Sn for heavier fissioning systems in the isotopic chain of uranium, for initial excitation energy around $\sim 1$MeV above the fission barrier. Similar values are also obtained using the GEF code. The data from both $\gamma$-spectroscopy measurements strongly disagree with these predictions, while the present data, together with previous direct isotopic measurements for the fission of $^{238}$U~\cite{Ramos2018,Pellereau2017}, follow the trend and values predicted by the models. 
The large deviation of the fission yield for Sn obtained from $\gamma$-$\gamma$ coincidences~\cite{Wilson2017} can be seen from Figure~\ref{fig:SnYield}. The fission yield  obtained from  single $\gamma$-spectroscopy~\cite{Fotiades2019} has also a large deviation, even if the reported value is closer to the present data and fission models.

The comparison performed in this work between the present data from neutron-transfer induced fission and data from neutron-capture induced fission might be affected by the different populations of angular momentum in both processes. Recent experimental results obtained from radiative neutron-capture processes and from surrogate reactions showed a strong enhancement of the $\gamma$-emission probability induced by the surrogate reaction with respect to the direct reaction~\cite{Boutoux2012}, which was attributed to a larger angular momentum populated in the former reaction. 
However, this behavior was not observed in fission, where neutron-induced and transfer-induced fission show similar probabilities~\cite{Kessedjian2010} and fission-fragment distributions~\cite{Ramos2018}. Small variations due to angular momentum were also predicted by theoretical models. The \textit{Metropolis walk} method combined with shape evolution, based on microscopically calculated level densities~\cite{Ward2017}, predicts a negligible variation in yields of the atomic number for angular momenta above $J=2\;\hbar$. The GEF code~\cite{GEF} estimates a variation lower than $5\;\%$ in the yield of Sn for angular momenta ranging from $J=3-10\;\hbar$. These observations, together with model predictions, exclude the influence of angular momentum in the discussion about the discrepancies in Mo and Sn between the present data and data from $\gamma$-spectroscopy~\cite{Wilson2017,Fotiades2019}.

The discrepancies between direct and $\gamma$-spectroscopy measurements may partially result from the bias arising from the $\gamma$-ray multiplicity experimental selection, as discussed in Ref.~\cite{Fotiades2019} and pointed out in the theoretical work of Ref.~\cite{Pasca2018}. Further, the  $\gamma$-spectroscopy measurements rely on the precise knowledge of long-lived isomeric states that are known to occur in the  Sn region (for instance the $7^{-}$ state with a half-life of the order of milliseconds in $^{128}$Sn and $^{130}$Sn~\cite{Fogelberg1981}). All these transitions need to be considered to determine accurately the fission yields from $\gamma$-spectroscopy measurements. However, the knowledge of these transitions is far from being exhaustive today, as new transitions are still being found~\cite{Pietri2011,Simpson2014} and fission yields extracted using $\gamma$-spectroscopy  could be  underestimated. This further highlights the importance of direct and complete isotopic fission-fragment yields to obtain an accurate modeling of the fission process. 

In summary, the first direct measurement of isotopic fission yields of $^{239}$U, performed using the neutron-transfer $^{9}$Be$(^{238}$U,$^{239}$U$)^{8}$Be reaction, is reported. An overall agreement on the fragment yields and N/Z was achieved between the data and GEF calculations within 10\% of accuracy. The increase of the production yield of Sn with respect to the fissioning mass was experimentally found in $^{238}$U and $^{239}$U, in agreement with the predictions of the Brownian shape-motion model and GEF. The present results disprove the fission yield anomaly for Mo and Sn isotopes reported from recent indirect $\gamma$-spectroscopy measurements. The survival of the asymmetric \textit{Standard I} fission mode at excitation energies 2~MeV above the fission barrier was confirmed for $^{239}$U. This unique and complete data set provides reliable constrains for fundamental fission models and valuable inputs for the evaluation of the heating inside of current reactors and the incineration capabilities of future reactors. 

The authors acknowledge the excellent support from the GANIL technical staff, J. Goupil, G. Fremont, L. Menager, J.A. Ropert, and C. Spitaels. A. Navin is acknowledged for a careful reading of the manuscript. D.R. thanks PACS group (IPN Orsay) for the support and fruitful discussions. This work was partially supported by the Spanish Ministry of Research and Innovation under the budget items FPA2015-71690-P and RYC-2012-11585, and by the Xunta de Galicia under item ED431F 2016/002. The research leading to these results has received funding from the European Union's HORIZON2020 Program under grant agreement n$^{\circ}$654002.

\Urlmuskip=0mu plus 1mu\relax
\bibliography{e753_paper} 
\bibliographystyle{myapsrev4-1}
\end{document}